\documentclass[11pt,preprint]{aastex}
\usepackage{bm}

\begin{document}

\title{Implementation of robust image artifact removal in \textsc{SWarp}\\ through clipped mean stacking}
\author{D. Gruen\altaffilmark{1,2}, S. Seitz\altaffilmark{1,2}, G. M. Bernstein\altaffilmark{3}}
\altaffiltext{1}{University Observatory Munich, Scheinerstrasse 1, 81679 Munich, Germany}
\altaffiltext{2}{Max Planck Institute for Extraterrestrial Physics, Giessenbachstrasse, 85748 Garching, Germany}
\altaffiltext{3}{Department of Physics and Astronomy, University of Pennsylvania, 209 South 33rd Street, Philadelphia, PA 19104, USA}
\email{dgruen@usm.uni-muenchen.de}

\begin{abstract}
We implement an algorithm for detecting and removing artifacts from astronomical images by means of outlier rejection during stacking. Our method is capable of addressing both small, highly significant artifacts such as cosmic rays and, by applying a filtering technique to generate single frame masks, larger area but lower surface brightness features such as secondary (ghost) images of bright stars. In contrast to the common method of building a median stack, the clipped or outlier-filtered mean stacked point-spread function (PSF) is a linear combination of the single frame PSFs as long as the latter are moderately homogeneous, a property of great importance for weak lensing shape measurement or model fitting photometry. In addition, it has superior noise properties, allowing a significant reduction in exposure time compared to median stacking. We make publicly available a modified version of \textsc{SWarp} that implements clipped mean stacking and software to generate single frame masks from the list of outlier pixels.
\end{abstract}


\section{Introduction}

The creation of stacked images out of multiple single exposures of the same part of the sky is a common problem in astronomy. Important operations, such as object detection or shape measurement, require the combined information to make full use of the individual frame data. In addition, stacking provides the advantage of greatly reducing the amount of data that needs to be analyzed downstream, for many purposes without loss of information. 

Usually, images are combined by re-sampling each single frame to a common grid and then taking either the (inverse-variance weighted) mean or median of the individual pixels as the stacked value. Both these methods are implemented, for example, in the most commonly used stacking software \textsc{SWarp}\footnote{\texttt{http://www.astromatic.net/software/swarp}} (\citealt{2002ASPC..281..228B}, for examples of recent larger applications, see \citealt{2007MNRAS.375..213W,2008PASP..120..212G,2008SPIE.7016E..17M}). While the inverse-variance weighted mean is statistically optimal in the sense of minimal noise in the resulting stack, it requires a highly complete algorithm for rejecting regions in the individual frames containing artifacts, such as 
\begin{itemize}
\item \emph{cosmic rays} (CRs), i.e. energetic particles of cosmic or terrestrial origin that hit the detector and cause highly elevated counts in single or few neighboring pixels; the pattern of elevated counts on the CCD they produce depends on the particle and properties of the camera, with charge deposits varying typically between $~25-75$ electron-hole pairs per $\mu$m of the track \citep{2002ExA....14...45G}; they can be detected by their peaked profile that is incompatible with the extended point spread function of sources imaged by the atmosphere (e.g. \citealt{2000PASP..112..703R} or \citealt{2002AuA...381.1095G} for a method based on detecting features of higher spatial frequency than the PSF, \citealt{1995PASP..107..279S} for a machine learning approach or \citealt{2001PASP..113.1420V} for a method based on edge detection),
\item \emph{tracks} of satellites, meteors, planes and sufficiently rapidly moving objects (e.g. comets and asteroids, which appear at different positions in different exposures) in the solar system are unwanted features in extra-solar astronomy; one method of detecting tracks for automated masking is based on the Hough transform \citep[e.g. in the implementation of][]{2001misk.conf..595V},
\item \emph{ghost images}, which appear as diffuse secondary images of bright stars and can be masked automatically only if accurate knowledge about reflecting surfaces and positions of bright stars is available; charge persistence of bright stars causes similar problems and can depend on several factors \citep{2012SPIE.8446E..3PG}, and
\item \emph{CCD imperfections}, such as saturation features, bad pixels or flat-fielding non-uniformities; these are most often stationary between different exposures and can therefore be taken care of with an instrument mask are easy to detect on single frames (e.g., by masking pixels above a saturation threshold), which is why we are not concerned with them in this work.
\end{itemize}
One major concern is that no algorithm working on the single frames is easily able to detect and correct all of these features that span a wide range in size and apparent surface brightness, while each of them can disturb photometric or  shape measurements significantly and, more importantly, potentially cause correlated errors on a certain scale and/or in a certain region of the sky. Mean stacks will, unless these features are detected and masked at the single-frame level, contain a density of artifacts that increases with the number of frames and total exposure time entering the stack. This has led many studies in the past to consider median stacks instead, which are more robust to single outlier frames when a large enough number of overlapping exposures is available.

Median stacks, however, are problematic in two important respects as well. For large, outlier-free samples, the variance of the median is a factor of $\pi/2\approx1.57$ larger than the variance of the mean, such that the total exposure time must be increased by more than half if stack images of similar depth are to be produced. They will, however, even then be far from optimal for a variety of analyses for the following reason. If the point-spread function (PSF) is not constant over the set of images entering the stack, the resulting stellar images will be a non-linear combination of the individual PSF profiles, piecewise equal to the individual profiles. The transformation of true surface brightness to observed surface brightness will be different for any object, and will no longer be described by a simple convolution, which renders median stacks almost useless for weak lensing shape measurement purposes. For photometry, this also means that PSF-convolved model fitting is not accurate and that the median stack is not necessarily flux conserving.

As an alternative, \textit{clipped mean stacks} have been created before using, for instance, \textsc{IRAF},\footnote{\texttt{http://iraf.noao.edu/}} \textsc{Drizzle} \citep{2002PASP..114..144F} and related software, \textsc{THELI} \citep{2005AN....326..432E}, the \textsc{O.A.R. IDL Library}\footnote{\texttt{http://www.oa-roma.inaf.it/}} or non-public software \citep[cf., e.g.,][]{2011arXiv1111.6619A,2012ApJ...761...15L,CFHTLST0007,2012AuA...544A.156M}. 

Three issues remain to be addressed, however. Firstly, clipped mean stacking is not presently implemented in \textsc{SWarp} despite its widespread use. We therefore make publicly available a modification of \textsc{SWarp} that implements this. Secondly, information about outliers with respect to the stack can be used in generating masks for the single frame images included. This is of great value for analyses making use of the set of individual exposures instead of the stack, as is the case for instance for upcoming shear measurement codes. Moreover, the spatial density of outliers can also be used to mask even lower significance features on the single frame level than is possible with clipping alone, and we release a piece of software that implements this. The result, which we call \emph{outlier filtered stack}, is cleaned even of faint tracks and ghost images, and it also provides cleaned single frames useful, for instance, for lensing purposes. Thirdly, we investigate the influence of differences in PSFs between the single frames entering the stack on the performance of these methods. We find that a simple clipped mean can distort the profiles of bright stars even at moderate thresholds. We suggest both a means of mitigating the effect and a metric for determining the required level of allowance for PSF differences.

In Section~2, we detail the algorithms used. In Section~3, we study the properties of the noise in these stacks and analyse the tolerance of the method for PSF inhomogeneity. Section~4 shows examples of our application of the algorithm to a set of real and simulated images. Section~5 provides practical information for downloading and applying the provided software. Section~6 summarizes our findings.

\section{Method}
\label{sec:method}
Consider a set of $M$ overlapping single frames $\mathfrak{F}=\lbrace F_1,F_2,\ldots,F_M\rbrace$. For every position in the overlapping region, we can interpolate each of the images to get the corresponding surface brightnesses at that point, $\mathfrak{f}=\lbrace f_1,f_2,\ldots,f_M\rbrace$. The median of $\mathfrak{f}$ is denoted as $\mu$. Let all frames be background subtracted, i.e. $\mathbb{E}(f_i)=0$ for pixels not containing astronomical objects. If all frames were on the exact same grid and shared the identical point-spread function, then $\mathfrak{f}$ would be a set of independent measurements of the same (convolved) true surface brightness at that point in the sky. Their uncertainties at the given point are denoted as $\lbrace \sigma_1, \sigma_2,\ldots,\sigma_M\rbrace$.

The uncertainty in a given pixel can be written as the sum of several components,
\begin{equation}
\sigma_i^2=f_{\rm obj}/g+f_{\rm bg}/g+\sigma^2_{\rm rn} \; ,
\end{equation}
where $f_{\rm bg}$ is the background flux, $f_{\rm obj}$ the object flux, both given in analog-digital units (ADU) with gain $g$, and $\sigma_{\rm rn}$ the read noise (all of which can have pixel-to-pixel variations). The $\sigma_i$ can, in principle, be taken directly from the weight frames, if these include all components and, in addition, are re-scaled correctly with the respective science images. Neither of this is reliably the case in pipelines in use today. Weight frames are often used merely as \emph{relative} weights
\begin{equation}
w_i=a\times \sigma^{-2}_{\rm bg} \; ,
\label{eqn:wa}
\end{equation}
inversely proportional to the background noise $\sigma^2_{\rm bg}=f_{\rm bg}/g+\sigma^2_{\rm rn}$ with arbitrary scaling $a$ (for a counter example of a full treatment of noise, see \citealt{2002AuA...381.1095G}).\footnote{Note that the background noise only weight has important advantages, since weighted mean stacking according to the full pixel noise $\sigma_i$ would distort the profiles of bright objects when frames have inhomogeneous PSFs. Object flux uncertainties, on the other hand, can easily be defined to include the photon noise of the object itself, as it is done in \textsc{SExtractor}.} 

\textsc{SWarp}, which we use as a basis for our implementation, allows various definitions. In the most common relative inverse-variance weight frame format \texttt{MAP\_WEIGHT}, it determines the scaling factor $a$ in Eqn.~\ref{eqn:wa} by matching $w_i^{-1}$ to the empirical variance in empty sky region of the single frames (found by means of $\sigma$-clipping to remove object pixels and subtraction of a background map). In addition, \textsc{SWarp} keeps track of the gain in the individual re-scaled single frames. Therefore, we can calculate the full pixel noise as
\begin{equation}
\sigma_i^2=\mu/g+\sigma^2_{\rm bg} \; ,
\label{eqn:pixelnoise}
\end{equation}
assuming that the median is an artifact-free model for the true surface brightness.

True astronomical single frame images of the same portion of the sky are different in at least four ways beyond having different realizations of noise:
\begin{itemize}
\item No two single frames are sampled on exactly the same grid. The astrometric solution allows to map pixel coordinates to common coordinates, and the imperfections due to the interpolation procedure are small.
\item No two PSFs are the same, at least in ground-based imaging. As a result of this, the \emph{measured} surface brightnesses are different in each of the images because the light is spread out differently. This effect is most severe for point sources. When the ratio of the widths of two PSFs is $\nu$ then, given flux normalization, their central amplitudes scale as $\nu^{-2}$.
\item There are additional effects of astronomical and terrestrial origin (listed in the Introduction and called artifacts in the following) that are different in each of the single frames; these are exactly what we are trying to detect and mask.
\item Variable or transient objects, which are likely to be interpreted by our algorithm as artifacts and clipped as well. Depending on the scientific goal, this can be either a good or an adverse effect.
\end{itemize}
If there were no difference in PSFs, we could perform a simple $\kappa-\sigma$ clipping to detect points that are influenced significantly by an artifact. Since $\mathfrak{F}$ has a certain distribution of PSFs, however, we will have to allow for PSF related scaling of the flux.

\begin{figure}[!h]
\epsscale{0.45}
\plotone{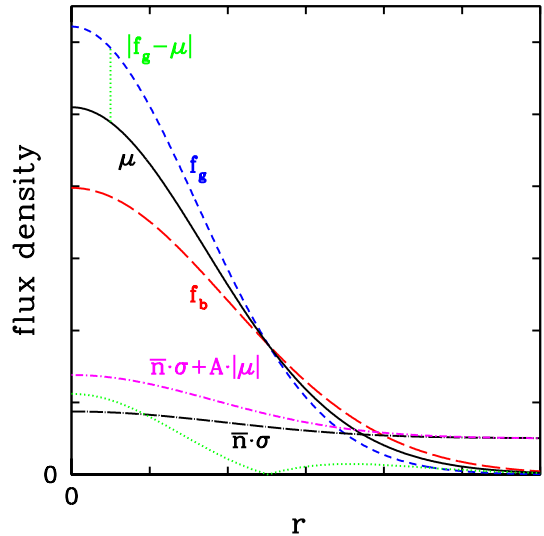}
\caption{Schematic picture of the clipping procedure in the presence of inhomogeneous PSFs. Given a set of frames with varying seeing, the photometrically scaled profile of a point source varies between $f_g$ (\textit{blue, short dashed line}) at good and $f_b$ (\textit{red, long dashed line}) at bad seeing conditions. The median profile is $\mu$ (\textit{black, solid line}). The difference between the median profile and the best seeing profile, $|f_g-\mu|$ (\textit{green, dotted line}) exceeds the threshold $\bar{n}\cdot\sigma$ (\textit{black, dotted-long-dashed line}) in the inner part. When increasing the threshold to $\bar{n}\cdot\sigma+A\cdot|\mu|$ (\textit{magenta, dotted-dashed line}), however, this can be compensated and the clipping of artifact-free stars is avoided.}
\label{fig:sketch}
\end{figure}

The following criterion takes this into account so as to robustly clip true outlier pixels. Let $\mu$ be the median of $\mathfrak{f}$. The value $f_i$ is rejected in our algorithm if
\begin{equation}
|f_i-\mu| > \bar{n}\sigma_i+A|\mu| \;,
\label{eqn:clip}
\end{equation}
where $\bar{n}$ and $A$ are two parameters of the clipping specifying the statistical and PSF related leniency of the procedure. Recall that the frames are background-subtracted, i.e. $\mu\approx0$ in empty sky. The choice of $A$ will depend on the distribution of PSFs and the choice of $\bar{n}$ on the particular purpose, discussed in more detail in Sections 2.1 and 3.2. For $A=0$, this produces what is simply a weighted mean stack clipped relative to the median. An exemplary sketch of why $A>0$ is usually required in the face of PSF inhomogeneity is shown in Figure~\ref{fig:sketch}.

This procedure works best in the limit of a noiseless median, i.e. with a large number of exposures. We propose the following rule in the case where only two frames are available, i.e. $\mathfrak{f}=\lbrace f_1, f_2\rbrace$: if
\begin{equation}
|f_1-f_2| > \bar{n}\sqrt{\sigma_1^2+\sigma_2^2}+A\left|\frac{f_1+f_2}{2}\right| 
\end{equation}
is true, we discard the pixel information from both frames. The reliability of the algorithm, however, remains strongly dependent on the number of available exposures.

\subsection{Spatial filtering of outliers}

In order not to remove too much information from the image, it is necessary to use a threshold $\bar{n}$ large enough such that most of the $f_i$ enter the weighted mean, i.e. the stacked output value (cf. also Section~3.1). Many artifacts, however, are at a surface brightness level which is comparable to the sky noise in a single frame. They obviously cannot be removed reliably on the basis of single pixel clipping. They introduce, however, a higher spatial density of outlier pixels at the artifact position. A first run of the clipped mean stacking with a relatively low threshold can be used to generate a list of outlier pixels. Their density can be used to mask artifacts on the single frame level. A final run which stacks the masked single frames in a mean stack produces an outlier filtered stack with optimal noise properties. This section is concerned with the statistical aspects of such a scheme.

\begin{figure}
\centering
\epsscale{0.45}
\plotone{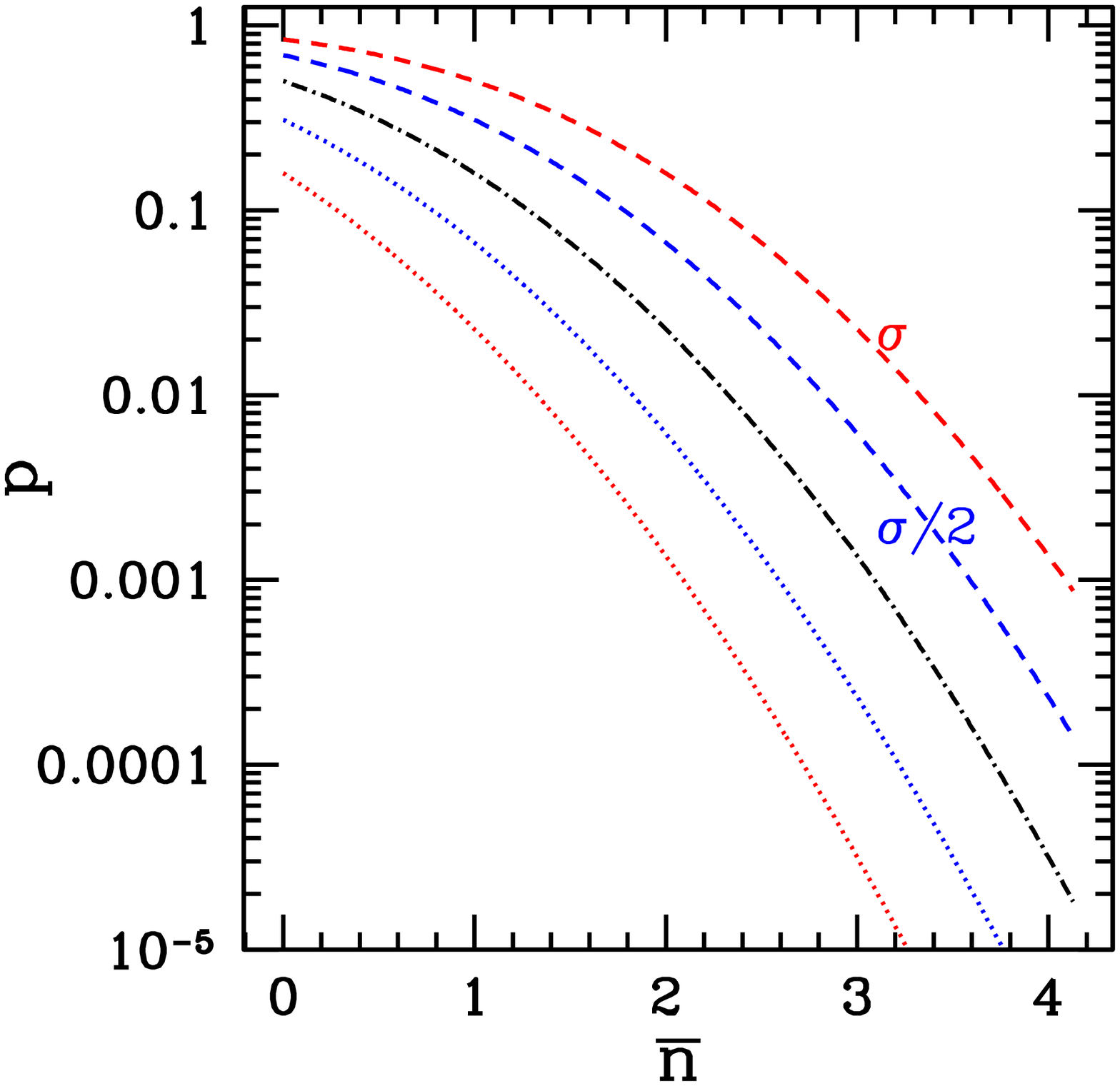}
\caption{Outlier probabilities for Gaussian noise. Plotted are the probability $\hat{p}=\check{p}$ for any pixel to be more than $\bar{n}\cdot\sigma$ above (or below) the true value for a correct frame (\textit{black, dotted-dashed line}) and the corresponding probabilities $\hat{p}$ and $\check{p}$ for being above (\textit{dashed lines}) or below (\textit{dotted lines}) if there is an artifact with positive surface brightness of $n\cdot\sigma$ (\textit{blue, closer to central line:} $n=0.5$; \textit{red, further from central line:} $n=1$) present in the pixel.}
\label{fig:gaussout}
\end{figure}

Consider a Gaussian distribution of noise with zero mean and some standard deviation $\sigma>0$, which determines the statistical deviation of a pixel from the true surface brightness at the corresponding point in the sky.\footnote{The noise is in fact Poissonian, but at large enough background levels as they are common in optical imaging this can safely be approximated by a Gaussian distribution.} We can count pixels that are outliers by selecting those which are $\bar{n}\cdot\sigma$ or more off (above or below) the expected true surface brightness at the respective point. Due to random fluctuations, this will flag a certain fraction of pixels. Figure~\ref{fig:gaussout} shows the probability $\hat{p}$ of a pixel being more than $\bar{n}\cdot\sigma$ above the true value (dotted-dashed black line), which is related to the Gaussian error function,
\begin{equation}
\mathrm{erf}(x)=\frac{2}{\sqrt{\pi}}\int_0^x e^{-z^2}\; dz
\end{equation}
as
\begin{equation}
\hat{p}=\frac{1-\mathrm{erf}(\bar{n}/\sqrt{2})}{2} \; .
\end{equation}
With no artifacts present, $\hat{p}$ equals the probability $\check{p}$ of being as far \emph{below} the true value as well.

Now consider an artifact in one frame that raises the surface brightness in some region by $n\cdot\sigma$.\footnote{Note that for an artifact of given physical surface brightness, $n$ depends on the noise level at the particular position. This corresponds to the fact that an artifact of fixed surface brightness will be more difficult to detect in regions where background noise or shot noise from bright objects are larger.} The probability of high (low) outliers with respect to the true surface brightness will then increase (decrease) for any threshold $\bar{n}$. Figure~\ref{fig:gaussout} shows the increased per-pixel probability of values more than $\bar{n}\cdot\sigma$ above the mean as the dashed curves, which are plotted for $n=+0.5$ and $n=+1$.

In the case of practical application, two complications arise. For one thing, the true surface brightness in a pixel is not known. While the median of a sufficiently large number of frames is a robust estimate, its uncertainty adds to the intrinsic scatter of counts in single-frame pixels, for which some allowance must be made. Secondly, the counts of outlier pixels are increased when inhomogeneity in the PSF profile is present (see also Section~3.2). We therefore use the same argument as in Eqn.~\ref{eqn:clip} to define
\begin{equation}
n=\mathrm{sgn}(f_i-\mu)\cdot\frac{\mathrm{max}(|f_i-\mu|-A|\mu|,0)}{\sigma_i}
\label{eqn:n}
\end{equation}
as the outlier significance.

Our concept for masking diffuse artifacts therefore is the following. Since outliers are more likely in an area with systematic surface brightness offsets (even though not all or potentially not even a majority of pixels will be above a sensible threshold) we can find a mask by selecting areas, for instance squares of area $N$ pixels, inside which more than $\bar{N}$ pixels are more than $n>\bar{n}$ off the expected (median) value. Note that in this procedure we discriminate counts of positive and negative outliers, since they are more significant indicators of artifact flux or flux decrement when used individually rather than added. The question which thresholds and mask sizes should be used is discussed further in the following and in Section~4.

\subsubsection{Statistics of Outlier Counts}
\label{sec:statout}
In our simple outlier filtering algorithm, an area of $N$ pixels shall be masked if more than $\bar{N}$ pixels are (without loss of generality) positive outliers by more than $\bar{n}$ standard deviations. The purpose of this section is to determine optimal thresholds $\bar{N}$ and $\bar{n}$ as a function of area size $N$, which allow the detection of low surface brightness features at a small false positive rate. Our simplifying assumption in this context is that the artifact surface brightness is a noiseless constant multiple of the pixel noise over the size of the mask. For an overview of the quantities used in this context, we refer the reader to Table~\ref{tbl:quantities}.

\begin{table}
\begin{center}
\begin{tabular}{|l|l|}
\hline
$n$ & number of standard deviations $\sigma$ by which a pixel deviates from the true surface brightness \\ \hline
$\bar{n}$ & threshold of $|n|$ above which a pixel is counted as an outlier \\ \hline
$N$ & number of pixels of a rectangle that is being tested for artifacts \\ \hline
$\bar{N}$ & threshold of the number of same-sided outlier pixels in a rectangle above which it is \\ & considered contaminated by artifacts and masked \\ \hline
$\hat{p}$ & $\hat{p}=\mathrm{Prob}(n>\bar{n})$ for one pixel and the selected threshold $\bar{n}$ \\ \hline
$\hat{P}$ & probability of finding more than $\bar{N}$ high outliers above $\bar{n}$ inside one rectangle of $N$ pixels \\ \hline
\end{tabular}
\end{center}
\caption{Reference of common symbols for Section~\ref{sec:statout}.}
\label{tbl:quantities}
\end{table}

The probability $\hat{P}$ of finding more than $\bar{N}$ high outliers in $N$ pixels can be calculated using the Binomial probability distribution,
\begin{equation}
\hat{P}=\sum_{m=\bar{N}}^{N} \mathcal{B}(m; N, \hat{p}) \; ,
\label{eqn:pbin}
\end{equation}
where
\begin{equation}
\mathcal{B}(m; N, \hat{p})=\left(\begin{array}{c}N\\m\end{array}\right)\cdot \hat{p}^m\cdot(1-\hat{p})^{N-m}
\label{eqn:binomial}
\end{equation}
is the probability of finding the sum $m$ of the results of $N$ Bernoulli experiments with success probability $\hat{p}$ (i.e. $m$ out of $N$ pixels as high outliers). For $N\cdot \hat{p}>10$ and $N\cdot(1-\hat{p})>10$ we approximate Eqn.~\ref{eqn:binomial} by a Normal distribution, including the continuity correction \citep{continuity}, as
\begin{equation}
\hat{P}\approx\left(1-\mathrm{erf}\left(\frac{\bar{N}-N\cdot \hat{p}-0.5}{\sqrt{2N\hat{p}(1-\hat{p})}}\right)\right)/2\; .
\end{equation}
For the other limiting case of $N>20$ and $\hat{p}<0.05$ we use the Poissonian approximation,
\begin{equation}
\mathcal{B}(m; N, \hat{p})\rightarrow\frac{\lambda^m\cdot e^{-\lambda}}{m!}\; ,
\end{equation}
with $\lambda=N\cdot\hat{p}$, which can be applied equivalently for $\hat{p}>0.95$ when replacing the event by the non-event accordingly.

We make an a priori choice of the mask size $N$ and the tolerable false-positive rate $\hat P_{\rm max}$ that we are willing to accept. The problem is then to find thresholds $\bar{N}$ and $\bar{n}$ for optimal detection of low surface brightness features under these conditions. This can be done in two steps, 
\begin{enumerate}
\item by finding the $\bar{N}$ as a function of $\bar{n}$ where the false positive rate is below $\hat P_{\rm max}$ and
\item by determining the surface brightness of outliers that could be detected as a function of threshold $\bar{n}$ when the above method of choosing $\bar{N}$ is applied (i.e., when the false positive rate is limited to a fixed value).
\end{enumerate}

\begin{figure}
\centering
\plotone{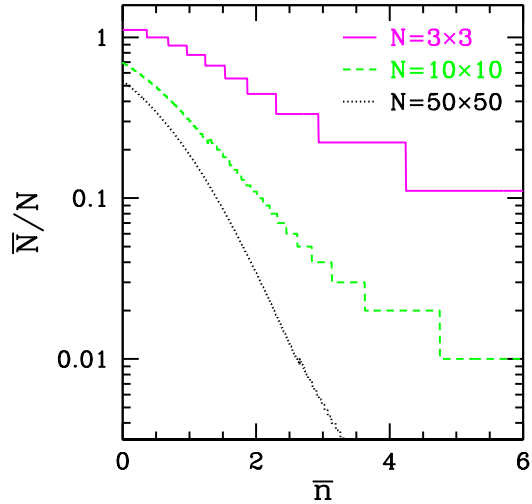}
\caption{Lowest $\bar{N}\in\mathbb{N}$ for which the probability of false detection of non-artifact area as an artifact is $\hat{P}<\hat P_{\rm max}=10^{-4}$ as a function of outlier threshold $\bar{n}$. With decreasing $\bar{n}$ and $N$ there is a strongly increasing chance of a large fraction of pixels being random outliers. For instance, for the smallest $3\times3$ box \emph{one} outlier pixel at $4.5\sigma$ significance is unlikely enough by chance ($\hat{P}<\hat P_{\rm max}$), but even \emph{all} pixels being above the mean ($\bar{n}=0$) is more likely than $\hat P_{\rm max}$.}
\label{fig:n04}
\end{figure}

The first step can be solved by iteratively applying Eqn.~\ref{eqn:pbin}. Figure~\ref{fig:n04} shows this for $\sqrt{N}=3,10,50$.

\begin{figure}[!h]
\centering
\plotone{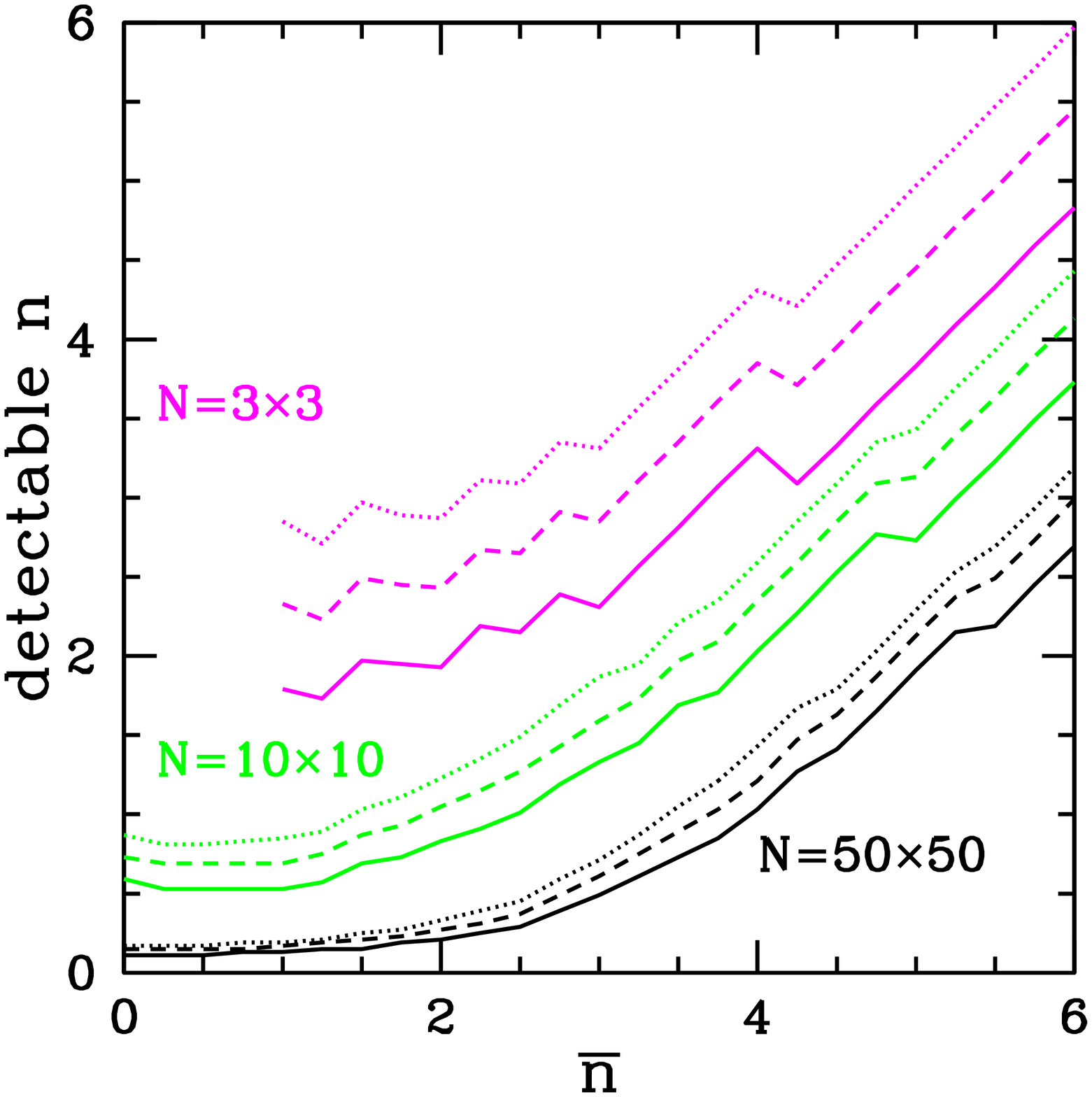}
\caption{Surface brightness level $n$ in units of sky noise detectable when filtering for outlier pixels in boxes of size $N=3\times3$~pixels (\textit{upper, magenta lines}), $10\times10$ (\textit{middle, green lines}) and $50\times50$ (\textit{lower, black lines}). Shown are the levels at which 68\% (solid lines), 95\% (dashed lines) and 99\% (dotted lines) of artifacts are detected using outlier counts above a threshold of $\bar{n}$, all at a false positive rate $\hat{P}<10^{-4}$ (cf. Fig~\ref{fig:n04}). For the $3\times3$ box, outliers of $2\sigma$ significance can be detected when using a threshold of $\bar{n}=2$. For the largest $50\times50$ box and a threshold of $\bar{n}=2.5$, features with a homogeneous surface brightness of only half the pixel noise can be detected in 99\% of the cases.}
\label{fig:nlim}
\end{figure}

For the second step, we again apply Eqn.~\ref{eqn:pbin} to find the surface brightness in units of the standard deviation as a function of $\bar{n}$ that are detected at $1\sigma$, $2\sigma$ and $3\sigma$ significance (i.e., in approximately 68\%, 95\% and 99\% of the cases) at the tolerated false positive rate. The result of this calculation is shown in Figure~\ref{fig:nlim}.

The curves in Figure~\ref{fig:nlim} have minima at low thresholds, which consequently yield the largest amount of information about potential artifacts. However, the loss of detection significance is relatively small when increasing $\bar{n}$ from, for instance, 1 to 2.5 (in the case of $N=50\times50$, for example, this increases the detectable artifact surface brightness level from approximately $0.25\sigma$ to $0.5\sigma$, which is still sufficiently low). This allows us to limit the analysis to the more significant outliers whose rareness speeds up the processing time of the procedure considerably.

In order to be sensitive to a range of scales, we recommend to use three $\sqrt{N}=3,10,50$.
\begin{itemize}
\item At $N=3\times3$, one can detect small outliers such as cosmic rays. These have high significance, may in the extreme case be limited to one pixel, however. For $\bar{N}=1$ and a corresponding $\bar{n}=4.5$ the false detection rate is below $10^{-4}$.
\item At $N=10\times10$ and $\bar{n}=2.5$, our false detection criterion is fullfilled at $\bar{N}=7$. This allows the detection of 95\% of homogeneous artifacts with a surface brightness of $1.2\sigma$ (cf. Fig.~\ref{fig:nlim}). The filter has the additional advantage that also thin linear features like tracks, for which $\bar{N}\approx\sqrt{N}$, are detected (although for successful detection they of course need to be of higher significance than homogeneous artifacts).
\item At $N=50\times50$, $\bar{n}=2.5$ and $\bar{N}=35$ yield a $2\sigma$ detection of large artifacts of a surface brightness $n=0.5$.
\end{itemize}
We note that these filters can be used independently, combining the resulting masks. Alternatively, it is possible to use one after another at increasing mask size, disregarding outliers which were masked by a smaller filter for the application of the larger one. Especially the criterion for $N=50\times50$ is very sensitive to low surface brightness features, and may be found to be masking large areas around linear features or regions with slight background subtraction offsets, in which case thresholds can be increased. When the PSF has significant differences between the single frames entering the stack, additional care must be taken to not suffer from clipping bright stars (cf. Section~3.2).

We release a piece of software that maps the outlier pixels to the single frame coordinate systems and generates masks by counting outliers in all square boxes of a set of sizes. The processor time for the generation of masks from the outlier list is less than one second for a 2k$\times$4k frame on a 2GHz core.

\section{Properties of the stack}

In this section we consider two properties of the stacked images created with the schemes described above. For clipped mean stacks, the pixel noise is always increased relative to a mean stack, which is discussed in Section~\ref{sec:noise}. Both the clipped mean stack and outlier filtered stack are tolerant only to some degree of difference between the PSFs of the individual frames, which we take into consideration in Section~\ref{sec:psf}.

\subsection{Noise}
\label{sec:noise}

\begin{figure}
\centering
\plotone{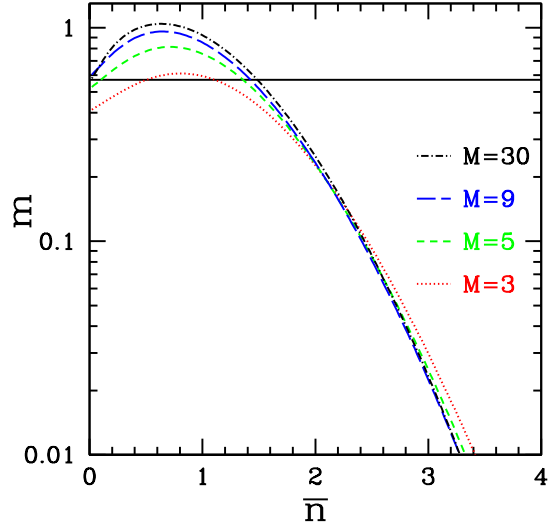}
\caption{Variance levels in images created with clipping at different numbers of overlapping frames $M=3,5,9,20$ as a function of clipping threshold $\bar{n}$. The excess variance $m$ is defined in Eqn.~\ref{eqn:excessvar}, with $m=0$ for a weighted mean and $m=1$ for a method that produces stacks of twice the pixel variance. All results are shown for $A=0.3$ using Eqn.~\ref{eqn:clip} and normalized by the variance of a mean stack. The solid horizontal line shows the noise level for a median stack in the limit of infinitely many exposures.}
\label{fig:noise}
\end{figure}

We study the level of noise in clipped mean images created using our algorithm as described in the first part Section~2. One can define a variance multiplier $m$ such that
\begin{equation}
\sigma^2(\bar{n})=(1+m(\bar{n}))\cdot\sigma^2 \; ,
\label{eqn:excessvar}
\end{equation}
where $\sigma^2$ is the sky variance of a mean stack, assuming Gaussian noise, and $\sigma^2(\bar{n})$ is the variance of a clipped mean stack with $\bar{n}$ and $A=0.3$ according to Eqn.~\ref{eqn:clip}. Plots for different numbers of overlapping frames are shown in Figure~\ref{fig:noise}.

We conclude that unless the clipping threshold is set below $\bar{n}<3$, the excess noise due to a simple clipping scheme is at or below the per cent level. Very low clipping thresholds should be avoided (except for the identification and masking of artifacts, cf. Section~2.1) and a median stack should be used instead. For the removal of low surface brightness features, we recommend the filtering scheme of Section~2.1. The weighted mean of single frames with artifacts masked by means of this allows for an outlier-free mean stack with optimal noise properties.

\subsection{Point-spread function}
\label{sec:psf}

The ideal stack has a PSF equal to a linear combination of the individual frame PSFs scaled according to their weight. This is what is the case for a mean stack, but it is not true for a median image. Here we discuss the influence of PSF inhomogeneity on the clipped mean or outlier filtered stack.

Let $s_{i,\bm{x}}$ denote the pixel at position $\bm{x}$ of the normalized PSF profile in a single frame $i=1\ldots M$. The median value in that pixel among the frames shall be $\mu_{\bm{x}}$. We can count the number of pixels in which the difference from the median is above the threshold,
\begin{equation}
\frac{\mathcal{F}\cdot\mathrm{max}(|s_{i,\bm{x}}-\mu_{\bm{x}}|-A|\mu_{\bm{x}}|,0)}{\sigma_i}>\bar{n} \; ,
\end{equation}
where we have calculated the deviation according to Eqn.~\ref{eqn:n} and scaled the normalized PSF by the total flux $\mathcal{F}$ of the brightest usable (i.e., non-saturated) stars. Where the count of either positive or negative outliers exceeds $\bar{N}$, a filter will clip a large fraction of bright stars. The resulting distortion of the PSF profile is problematic in similar ways as in a median stack, which is why it should be avoided.

\begin{figure}
\centering
\plotone{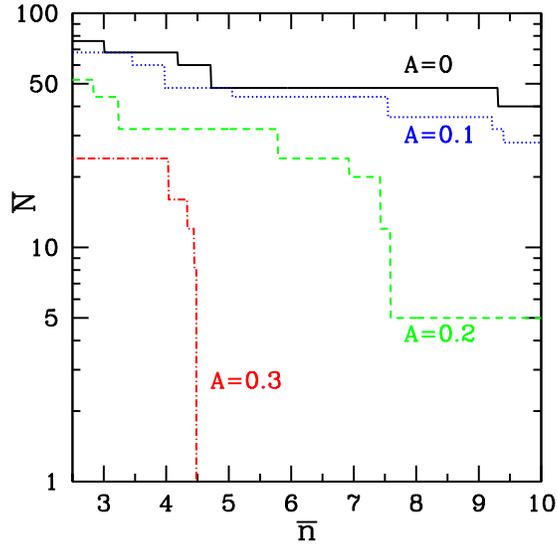}
\caption{Number of outliers expected for bright stars as a function of threshold $\bar{n}$. At the random distribution of PSF widths with $\sigma_{\log_{10} \rm{FWHM}}=0.05$ used here, only $A\geq0.3$ allows clipping of outliers at $\approx 5\sigma$ without a distortion of the stacked profile of bright stars. Even then, additional allowance for the number of $\leq4\sigma$ outliers has to be made for outlier filtering schemes.}
\label{fig:psf}
\end{figure}

We show results of calculating this metric for a set of 9 frames with circular Gaussian PSF of varying width. The widths are drawn from a log-normal distribution, centered on a FWHM of 4~pixels with $\sigma_{\log_{10} \rm{FWHM}}=0.05$ (approximately 10\%). This is an achievable configuration, although atmospheric seeing can potentially also vary considerably more. Figure~\ref{fig:psf} shows the maximum number of outliers $\bar{N}$ among all frames across a stellar image with flux $f=10^4\sigma$ as a function of threshold $\bar{n}$ for different parameters $A=0,\ldots,0.3$. Similar curves for a set of actual single frame PSFs can be calculated using the software \textsc{PSFHomTest} (see Section~5).

We find that at $A=0$, the common outlier clipping, the removal of even highly significant artifacts leads to a distortion of the PSF by clipping bright stars in the most extreme frames of the sample. In this case, a clipping scheme can only provide clean images with undisturbed PSF if high enough $A\geq0.3$ are used. Additionally, some allowance in the $\bar{N}$ thresholds chosen for outlier filtering must be made, since even then bright stars can lead to a large number of outliers with low individual significance.

We conclude that clipping schemes should only be applied when testing the homogeneity of the PSF and choosing parameters $A$, $\bar{n}$ and $\bar{N}$ for which the PSF is left undisturbed. A practicable way is to pick parameters $N=3\times3, 10\times10, 50\times50$, get $\bar{n}$ and $\bar{N}$ from Figures~\ref{fig:nlim} and \ref{fig:n04}. At what parameter $A$ these can be used without clipping bright stars can be tested with \textsc{PSFHomTest} (see Section~5). If the PSF inhomogeneity requires a large parameter $A>0.5$, it is recommendable to increase the thresholds $\bar{n}$ and $\bar{N}$ instead or split the set of frames in two or more seeing bins, as otherwise a high $A$ decreases the sensitivity to outliers blending with real objects. An alternative but computationally much more challenging approach would be the PSF homogenization or deconvolution of single frames before detecting outliers.

\section{Application to astronomical images}

\subsection{Simulations}

We simulate three main applications of the masking, namely cosmic ray, track and ghost image removal, to quantify the reliability of the algorithm in a more realistic manner. As a measure of efficiency, we use 
\begin{equation}
\eta=\frac{\mathrm{artifact\; flux\; masked}}{\mathrm{total\; artifact\; flux}} \; .
\end{equation}

\subsubsection{Cosmic rays}
\label{sec:cr}

The signature of cosmic rays on a CCD detector are most typically straight lines, while the counts and length depends on the particle energy and angle of entry. For a given energy spectrum, the mean length and count per pixel also vary with the thickness of the CCD \citep[cf.][]{2002ExA....14...45G}. We use a simple model to test the sensitivity of our algorithm, with a log-normal distribution of track length
\begin{equation}
p(\log [l/\mathrm{pix}])\propto \exp(-(\log [l/\mathrm{pix}] - 1)^2/(2\sigma_l^2))
\end{equation}
with $\sigma_l=0.3\approx\log(2)$. The width of the track is set to only one pixel. Assuming a deposit of 1000 charges per pixel \citep[cf.][]{2002ExA....14...45G}, we include shot noise in our simulations. We test the completeness of the masking with 9 simulated overlapping frames, a density of $2\cdot10^{-5}$~events/pix and varying levels of count deposit per pixel. 

\begin{figure}
\centering
\epsscale{0.48}
\plotone{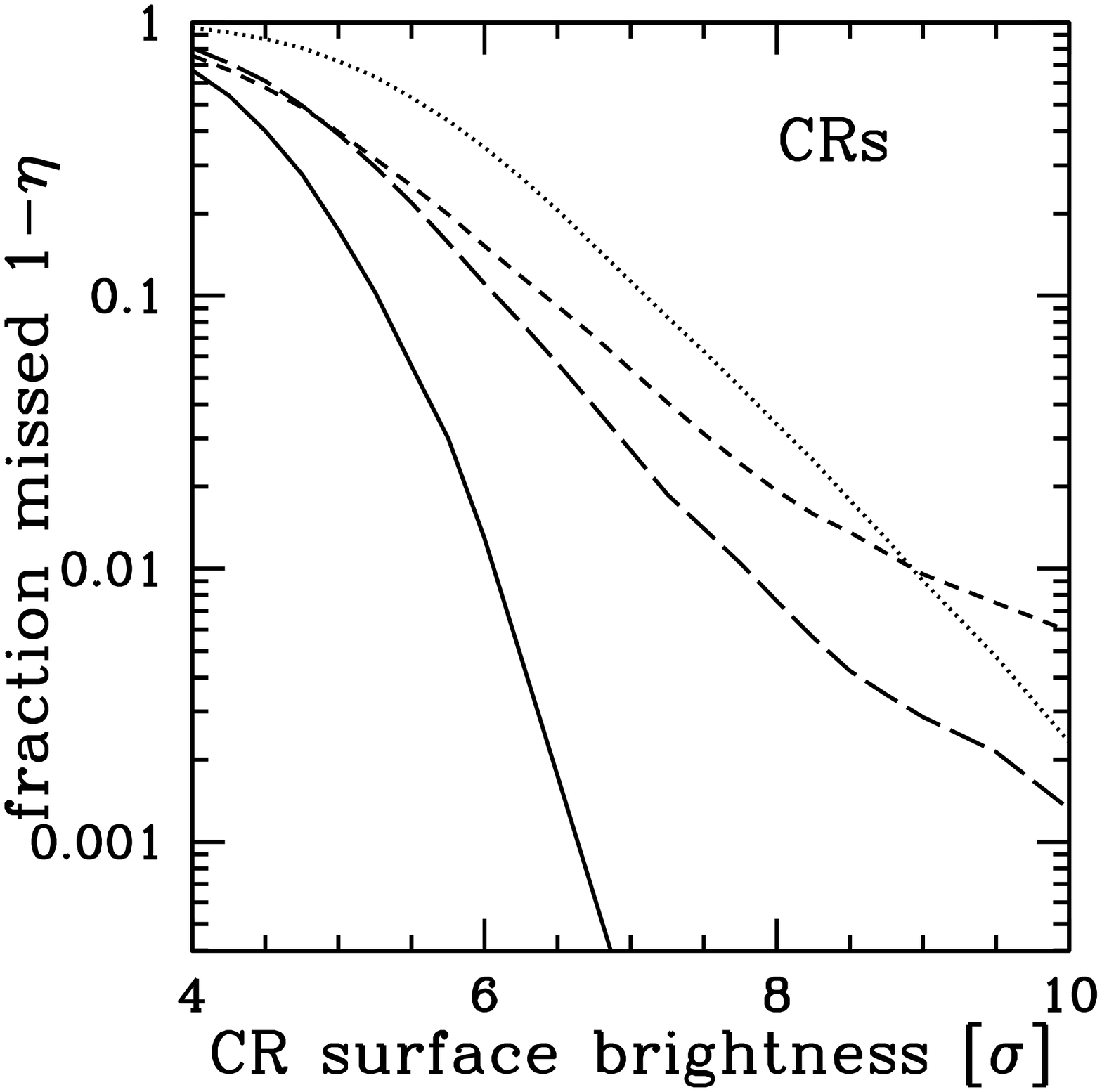}
\plotone{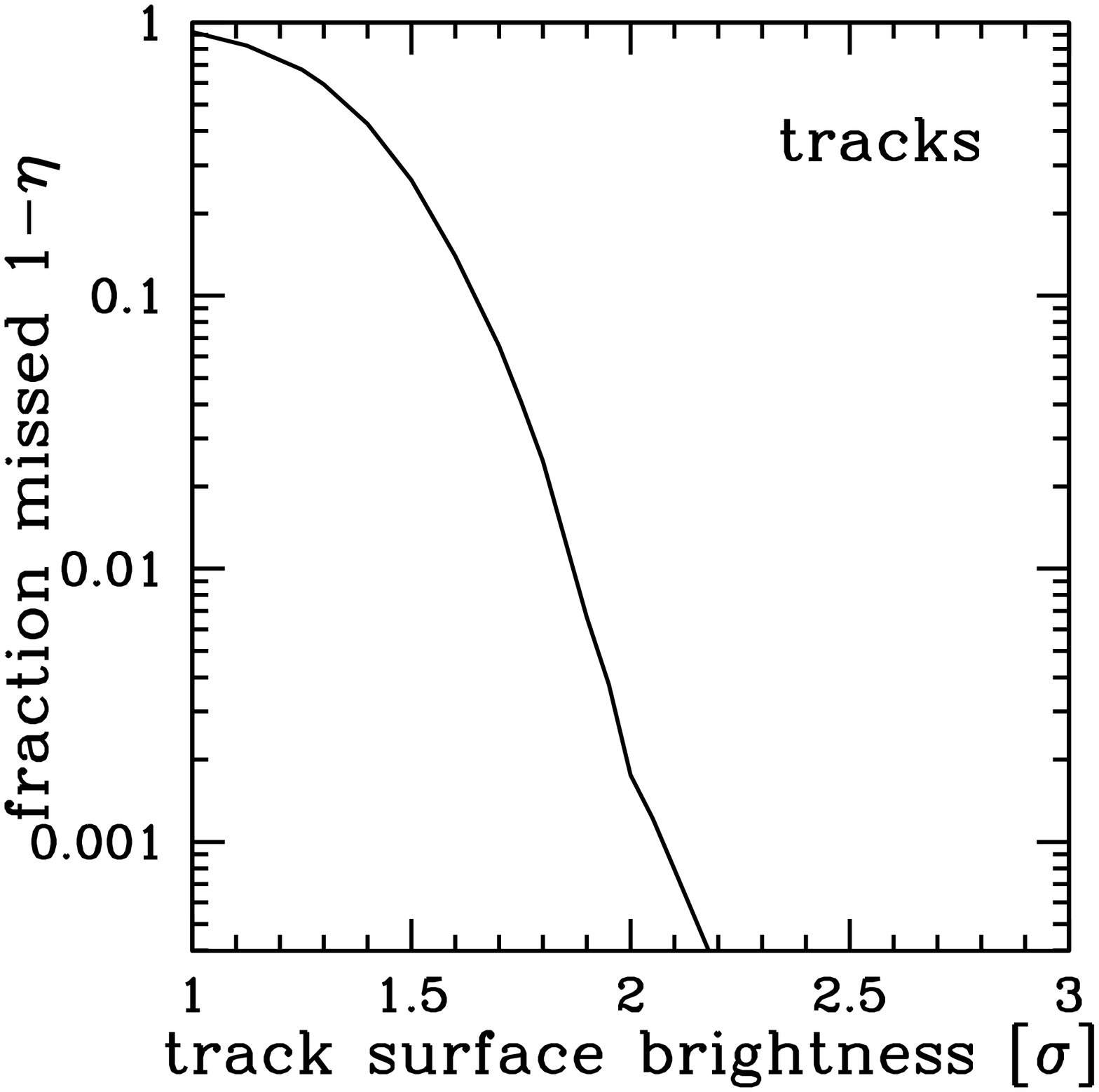}
\plotone{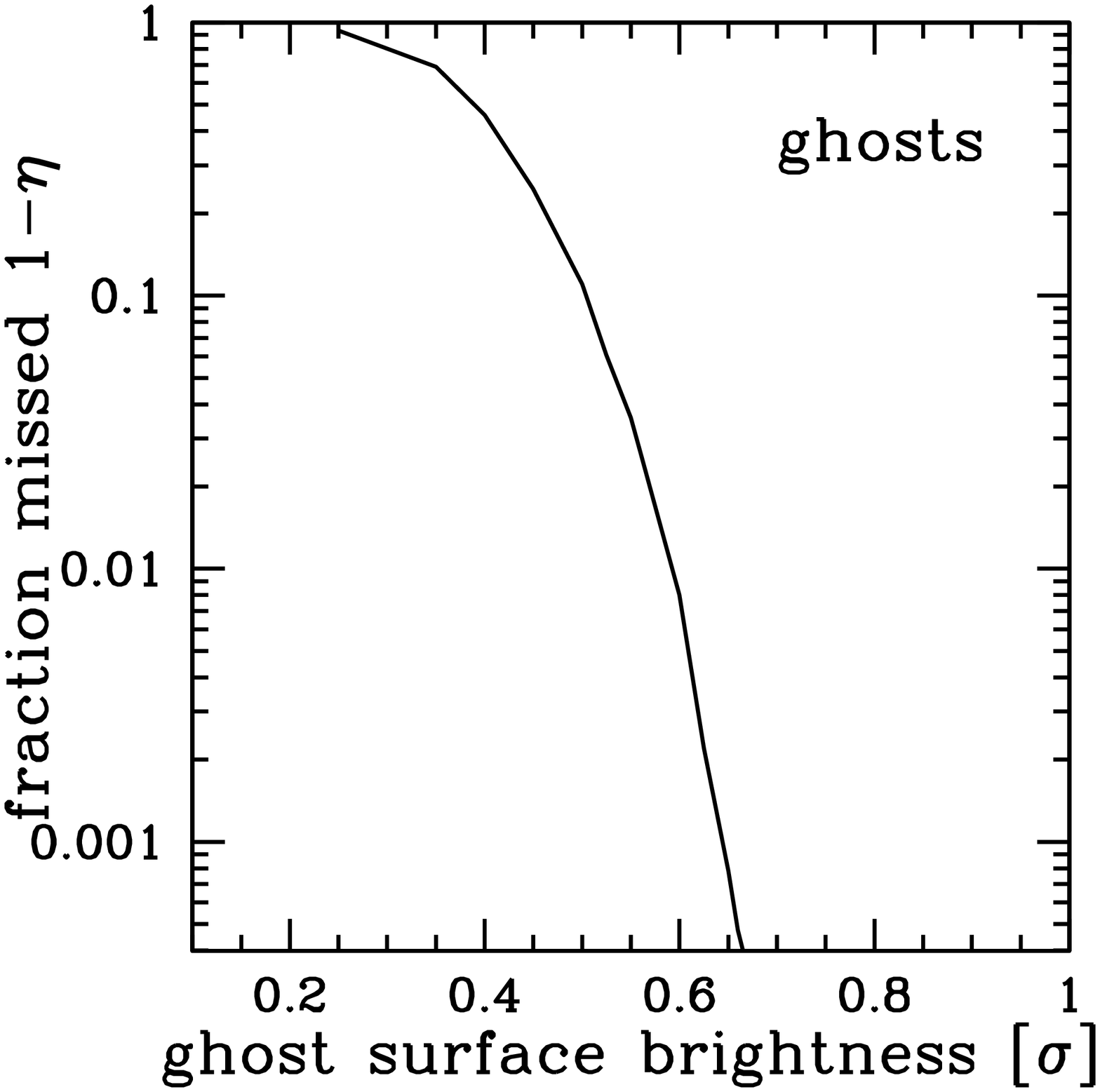}
\caption{Completeness of masking as a function of surface brightness in units of the sky noise for cosmic rays, tracks and ghost images. For cosmic rays, because they are most severely influenced by interpolation, we show different masking schemes: $N=3\times3$ with $\bar{n}=4.5$ and $\bar{N}=1$ (\textit{solid line}) on frames without interpolation, interpolated frames with the same settings (\textit{dotted line}) and with $\bar{n}=3$,$\bar{N}=2$ (\textit{short dashed line}), and interpolated frames with $N=4\times4$ boxes and $\bar{n}=3.5$,$\bar{N}=2$ (\textit{long dashed line}).}
\label{fig:cr}
\end{figure}

Because of the narrowness of the features, interpolation influences the detection probabilities most severely among the artifacts studied in this section. We therefore compare the efficiencies of different masking and interpolation schemes. For the baseline masking parameters, we choose a mask size of $N=3\times3$~pixels at a threshold of $\bar{n}=4.5$ and $\bar{N}=1$ with $A=0.3$. Without interpolation (or with nearest-neighbor interpolation, which does not dilute the artifact signal), the result is shown as the solid line in Fig.~\ref{fig:cr}. The other lines show the result for interpolated frames, where the cosmic ray flux is distributed over neighboring pixels. Consequently, completeness of the masking is significantly lower when using the same box size and thresholds as before (dotted line). Detection probabilities can be improved by demanding at least $\bar{N}=2$ outlier pixels either in a $N=3\times3$ box with $\bar{n}=3$ (short dashed) or in an $N=4\times4$ box with $\bar{n}=3.5$ (long dashed).

Note that cosmic rays in realistic images typically have an even higher significance than the regime probed here, typically above $10\sigma$, and will therefore be masked with almost perfect completeness in any of these schemes.

% \subsubsection{Tracks}

Moving objects in the sky such as satellites cause track-like features which are a nuisance to astronomy. We test our algorithm for track masking by simulating images with linear features of a Gaussian profile, where the surface brightness in a pixel at a separation $d$ from the line is set to
\begin{equation}
f(d)=n\cdot\exp(-d^2/(2\sigma_d^2)) \; ,
\end{equation}
where we use a full-width at half-maximum of 5~pixels and a truncation at $d>5$~pixels, corresponding to a hardly resolved object after convolution with a ground-based PSF.

We use a mask of $N=10\times10$~pixels with the settings $\bar{n}_0=2.5$, $\bar{N}_0=7$ and $A=0.3$. Results are shown in Figure~\ref{fig:cr}. At a track surface brightness of twice the sky noise, the masking becomes very efficient. Below this, a larger filter could still be used successfully (see Ghost image simulations), yet with a large area masked around the track.

\subsubsection{Ghost images}

Secondary (so-called ghost) images of bright stars come in a wide variety of forms, connected to the many different light paths possible in a complex optical system. We simulate a type of ghost image similar to the most common one in the WFI camera on the ESO/MPG 2.2m telescope \citep{1999Msngr..95...15B}.

We model these ghost images as an annulus between 50 and 100 pixels. The shape of the image is distorted to an ellipse with axis ratio $q$, uniformly distributed in $q\in[0.3,1]$, keeping the area constant. The ring is filled with constant surface brightness and offset far from the primary image. We simulate such features with a density of $1\cdot10^{-6}$ per pixel.

Figure~\ref{fig:cr} shows that a mask with size $N=50\times50$~pixels and thresholds of $\bar{n}_0=2.5$ and $\bar{N}_0=35$ with $A=0.3$ is capable of highly complete removal of ghost images, even when the surface brightness of the feature is only a fraction of the sky noise.

\subsection{Example images}

We show an example of a frame observed with the WFI camera on the ESO/MPG 2.2m telescope \citep{1999Msngr..95...15B}. Our sample contains 18 400~s and 600~s R band exposures of a field containing a bright star with a ghost image inside the field of view, all with a PSF FWHM below 0.9~arcsec (3.78~pixels). The single frames are reduced using the pipeline based on Astro-WISE\footnote{http://www.astro-wise.org/} \citep{2007ASPC..376..491V} as described in \citet{2013arXiv1304.0764G}, which masks cosmic rays with high confidence, tracks with a somewhat lower success rate and has in its standard form no ghost image masking technique available. We apply outlier masking with $A=0.3$ and three filter sizes of $N=3\times3$, $\bar{n}=5$, $\bar{N}=1$ for small artifacts, $N=10\times10$, $\bar{n}=3$, $\bar{N}=10$ and $N=50\times50$, $\bar{n}=2.5$, $\bar{N}=85$ for larger features, lenient enough for the small PSF variations in the sample.

\begin{figure}
\centering
\epsscale{0.85}
\plotone{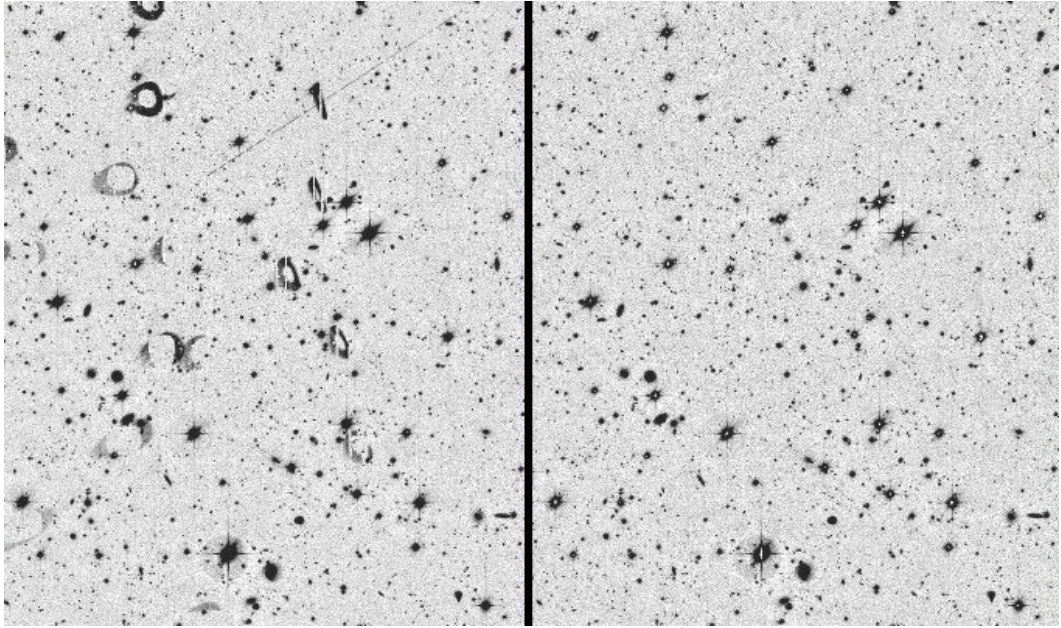}
\caption{Common mean stack (\textit{left panel}) and outlier filtered stack (\textit{right panel}) of a cutout from a field containing a bright star. Both ghost images and a satellite track are successfully removed from the image by outlier filtering.}
\label{fig:image}
\end{figure}

\begin{figure}
\centering
\epsscale{0.6}
\plotone{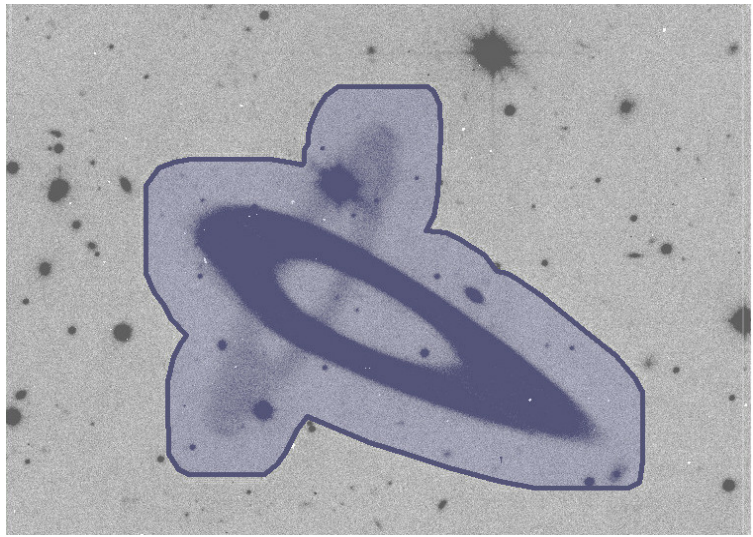}
\caption{Single frame with ghost image consisting of orthogonal bright and faint components. Since the surface brightness of the latter feature is less than $3\sigma$ above the background, a simple clipping cannot remove it reliably. The outlier filter mask (\textit{shaded region, edge marked as line for clarity}), however, detects and masks the feature entirely.}
\label{fig:mask}
\end{figure}

Figure~\ref{fig:image} shows a mean stack and an outlier filtered stack with the described settings. Both the multiple ghost images and a satellite track are successfully removed, while the depth of the image is conserved. The pupil ghost image close to saturated stars remains present since it does not change greatly between our exposures, which have only small dithers. Figure~\ref{fig:mask} shows a single frame image of a double ghost, overlaid with the mask generated with outlier filtering. Since the surface brightness of the fainter ghost is very low, only the combined information from a larger region is sufficient for detection and masking. The data cleaned from ghost images using the method described herein have been used successfully for cluster weak lensing analyses by \citet{wiscy}.

\section{Download and Application}

In this section we note some relevant details of our implementation of the described algorithm in \textsc{SWarp} and outline the procedure for practical application.

\subsection{Implementation}

\textsc{SWarp} applies a two-step forward mapping procedure to generate a stack out of single frames. In a first stage, images are background subtracted, resampled onto a common grid and saved temporarily. In a second stage, these images are combined into a stack according to a user selected \texttt{COMBINE\_TYPE} (such as \texttt{WEIGHTED} for weighted mean, \texttt{MEDIAN} or the \texttt{CLIPPED} mode implementing the algorithm described in this work). Both steps are frequently performed at different times, particularly in survey operations. This scheme implies that clipping can only happen at the second stage, where single frames are present in a comparable (background subtracted, flux scaled and re-sampled) form.

This also means that unlike in the implementation of clipping in \textsc{Drizzle} \citep{2002PASP..114..144F}, we cannot access the single frames on their original grid for comparison with a re-sampled median image, but must perform the clipping on re-sampled single frames. Consequently, it is more difficult to detect small features on the single frame which are smeared out by interpolation, although experience with cosmic rays at typical levels shows that they remain significant enough (cf. Section~\ref{sec:cr}). The scheme, however, also has the advantage not only of being more efficient computationally and from a pipeline perspective, but also of comparing two frames interpolated one time each instead of a non-interpolated and a doubly-interpolated version of the sky. Note that the masks found by means of outlier filtering are of course mapped back to the single frame grid straightforwardly.

One important issue is the correct treatment of pixel noise (cf. Section~\ref{sec:method}). When the weight map provided is based on background noise only (which is also true for weight maps generated by \textsc{SWarp} internally from the background level) and the individual frame gain keywords or the \texttt{GAIN\_DEFAULT} configuration parameter are set correctly, our calculation of pixel noise according to Eqn.~\ref{eqn:pixelnoise} works as intended. Alternatively, if the input weight map is based on the pixel noise including shot noise from object photons, gain should be set to 0, as otherwise the object contribution to the noise is added twice. Note that the latter setting is not recommended for the purpose of generating a weighted mean or clipped weighted mean stack, since it distorts surface brightness profiles in the presence of inhomogeneous PSFs (it can be used for outlier detection and masking on the single frame as described below, however).

\subsection{Practical use}

\subsubsection{Configuration}

For the basic configuration of \textsc{SWarp}, we refer the reader to the official documentation.\footnote{\texttt{http://www.astromatic.net/software/swarp}} The modifications of configuration parameters for clipped mean stacking are the following:
\begin{itemize}
\item \texttt{COMBINE\_TYPE} has a new option \texttt{CLIPPED}
\item \texttt{CLIP\_SIGMA} specifies the threshold parameter $\bar{n}$ from Eqn.~\ref{eqn:clip} (default: 4.0)
\item \texttt{CLIP\_AMPFRAC} specifies the parameter $A$ from Eqn.~\ref{eqn:clip} (default: 0.3)
\item \texttt{CLIP\_NAME} specifies the filename for the outlier list to be written (to be used for generating single frame masks, default: clipped.tab)
\end{itemize}

\subsubsection{Procedure}

We recommend the following procedure for generating clipped-mean or outlier filtered stacks. To remove highly significant outlier pixels in a clipped-mean stack,
\begin{enumerate}
\item generate single-frame PSF models with \textsc{PSFEx},
\item determine suitable clipping parameters with \textsc{PSFHomTest}, in particular make sure that the clipping threshold $\bar{n}$ and parameter $A$ do not clip pixels on bright stars, and
\item create a clipped mean stack with the modified version of \textsc{SWarp}.
\end{enumerate}
To produce an outlier filtered stack free of ghost images or faint tracks,
\begin{enumerate}
\item generate single-frame PSF models with \textsc{PSFEx},
\item determine suitable clipping parameters with \textsc{PSFHomTest}, in particular make sure that the combinations of $\bar{n}$ and $\bar{N}$ used for the filters later do not mask bright stars for the choice of $A$ taken,
\item create a clipped mean stack with the modified version of \textsc{SWarp} at a relatively low clipping threshold,
\item run \textsc{MaskMap} to generate single frame mask images from the outlier list,
\item multiply single frame weight images by the mask and
\item run \textsc{SWarp} again, yet in usual \texttt{WEIGHTED} mean mode.
\end{enumerate}

\textsc{PSFEx} and the regular version of \textsc{SWarp} are available from \texttt{http://www.astromatic.net}. The modified version of \textsc{SWarp}, \textsc{PSFHomTest} and \textsc{MaskMap} are available for download.\footnote{see \texttt{http://www.usm.uni-muenchen.de/\~{}dgruen/}} They include a C++ class for accessing \textsc{PSFEx} models, which might be useful for other purposes, too.

\section{Summary}

We presented a method of outlier rejection and filtering that successfully detects and masks unwanted features in astronomical images by comparison to the median stack. Simple outlier rejection removes highly significant outliers very efficiently. Calculations, simulations and practical application show that the outlier filtering method can be used to also mask lower surface brightness features such as tracks at more than twice and large area features such as ghost images at more than half the sky noise level above the background. It also has the benefit of generating single frame level masks that can be applied in analyses running on the single frame images.

One important caveat is that all clipping methods rely on some degree of homogeneity of the PSF. Simple outlier clipping changes the stacked profile of bright stars even at a relatively low level of PSF variation and moderate to high clipping thresholds. Differences in PSF profiles therefore must and can be accommodated, both for single pixel clipping and outlier filtering, but should be tested with the knowledge of single frame PSF models before the application of the scheme.

All software required and described in this paper is available for download (see Section~5).

\acknowledgments

This work was supported by SFB-Transregio 33 `The Dark Universe' by the Deutsche Forschungsgemeinschaft (DFG) and the DFG cluster of excellence `Origin and Structure of the Universe'.

We acknowledge the contribution of Emmanuel Bertin, the author of \textsc{SWarp}, which is a core component of the scheme presented here. The authors thank Arno Riffeser for helpful comments on the manuscript and Peter Melchior and Eric Suchyta for additional testing of the code.

\bibliographystyle{aamod_arno}
\bibliography{literature}

\end{document}